\documentclass[conference]{IEEEtran}
\usepackage{amsmath,amssymb,amsbsy,amsxtra,graphicx}
\usepackage{array,multirow}
\usepackage{booktabs}
\usepackage{url}
\usepackage{cite}
\usepackage{bm}
\usepackage{caption}
\usepackage{subfig}
\usepackage{epstopdf}
\usepackage{color}
\usepackage{balance}

\graphicspath{{./figures/}}

\begin{document}

\title{SignCol: Open-Source Software for Collecting Sign Language Gestures }


\author{\IEEEauthorblockN{Mohammad Eslami, Mahdi Karami, \\ Solale Tabarestani, Farah Torkamani-Azar}
\IEEEauthorblockA{DIgitial Signal Processing LAboratorY (DiSPLaY), \\Department of Electrical Engineering, \\ Shahid Beheshti University, Tehran, Iran\\
Email: \{m\_eslami, f-torkamani\}@sbu.ac.ir}
\and
\IEEEauthorblockN{Sedigheh Eslami, Christoph Meinel}
\IEEEauthorblockA{Department of Internet Technologies and Systems, \\ Hasso Plattner Institute, Potsdam, Germany\\
Email: \{sedigheh.eslami, christoph.meinel\}@hpi.de}
}

\maketitle

\begin{abstract}
Sign(ed) languages use gestures, such as hand or head movements, for communication. Sign language recognition is an assistive technology for individuals with hearing disability and its goal is to improve such individuals' life quality by facilitating their social involvement. Since sign languages are vastly varied in alphabets, as known as signs, a sign recognition software should be capable of handling eight different types of sign combinations, e.g.\ numbers, letters, words and sentences. Due to the intrinsic complexity and diversity of symbolic gestures, recognition algorithms need a comprehensive visual dataset to learn by. In this paper, we describe the design and implementation of a Microsoft Kinect-based open source software, called \textit{SignCol}, for capturing and saving the gestures used in sign languages. Our work supports a multi-language database and reports the recorded items statistics. SignCol can capture and store colored(RGB) frames, depth frames, infrared frames, body index frames, coordinate mapped color-body frames, skeleton information of each frame and camera parameters simultaneously. 


\end{abstract} 


\IEEEpeerreviewmaketitle

\section{Introduction}\label{Introduction}
It is predicted that by 2050, one in every ten people will have a severe hearing loss~\cite{WHO}. Absence of communication can have a significant impact on an individual's life, causing isolation and loneliness in society. World Health Organization (WHO) estimates that unaddressed hearing loss poses an annual global cost of 750 billion US\$  including loss of productivity for this population \cite{WHO}.

Sign language is the main communication way for people with hearing or speaking impairment. This type of language uses gestures as a manual way of communication to transfer a concept and a meaning to others. This can include simultaneously employing hand gestures, head actions, facial expressions, etc to convey a speaker's ideas. 

There is an undeniable communication problem between the people with hearing impairment and others. Innovations in automatic sign language recognition (SLR) is trying to tear down this communication barrier. 
Since sign languages have many aspects and details, SLR becomes a complicated visual task \cite{cheok2017review} and similar to any recognition task, a dataset is needed to be analyzed and used in training a SLR system \cite{baltruvsaitis2018multimodal,Mine,angra2017machine,tabarestani2015painting}. Due to the intrinsic complexity of the gestures used in sign languages, a comprehensive dataset is needed to cover the most essential signs for the recognition task. Researchers tried to collect their own dataset to validate their proposed approaches and methods for sign language recognition. Since collecting massive amounts of data is time-consuming and also there is not a unique protocol to collect the essential signs, the collected datasets are not comprehensive regarding sign variations and data modalities \cite{cheok2017review}. 

In this paper, we introduce \textit{SignCol (\underline{Sign Col}lector)}: an open-source software for collecting and storing gestures of sign languages. The main features and contributions of SignCol is summarized as follows \footnote{At the publication date, the authors, “Eslami” and “Tabarestani” are not at affiliated with Shahid Beheshti university anymore. They claim that there is no conflict of interest and all the content and software in this paper are based on the works which they had accomplished in Shahid Beheshti university}.

\begin{enumerate}
\item \textbf{Visual data capturing.} SignCol could connect to the Microsoft Kinect One (v2), capture and \emph{simultaneously} save different information and data modalities of the gestures, including: 
\begin{itemize}
\item RGB Color frames as images,
\item Infrared frames as images,
\item Depth frames as images,
\item Body index frames as images,
\item Skeleton information of frames as CSV files: the joint locations are in coordinates: 3D space of environment, 2D space of the color frame, 2D space of depth frame,
\item Coordinate mapped bodies (colorful bodies on depth space) as images,
\item All of the above data for the multi-person scene,
\item Parameters of the cameras from the Kinect as a text file.
\end{itemize}

\item \textbf{Database managing \& statistics reporting} SignCol includes database management for gathering, storing and reporting statistics. Number of defined items in each sign category and the number of captured items in each category are visualized in charts. This feature helps dataset collector to produce a uniform dataset of signs for different corresponding categories. In conclusion, user in SignCol is able to 
\begin{itemize}
\item define languages, e.g.\ English,
\item define sign items,
\item set the type of sign item in 8 categories (more in  \ref{overview}),
\item define performers,
\item consider the statistics of the dataset in charts and tables.  
\end{itemize}
\end{enumerate}

SignCol source files are shared online via github \cite{github}. Also, video demonstrations and binary execution files are available at \cite{video1}. The rest of this paper is organized as follow. Section \ref{background} describes background about sign languages, Microsoft Kinect and related work. The details about SignCol features, design and implementation and user interface forms are reported in section \ref{Main}. Finally, the paper is concluded in section \ref{conclusion}.

\section{Background}\label{background}

In this section, we provide more in-depth information
regarding Sign language and Microsoft Kinect. We also briefly introduce correctness rules for a few common rehabilitation exercises.

\subsection{Sign Language}\label{sign-language}

\begin{figure}[t]
\includegraphics[width=0.45\textwidth]{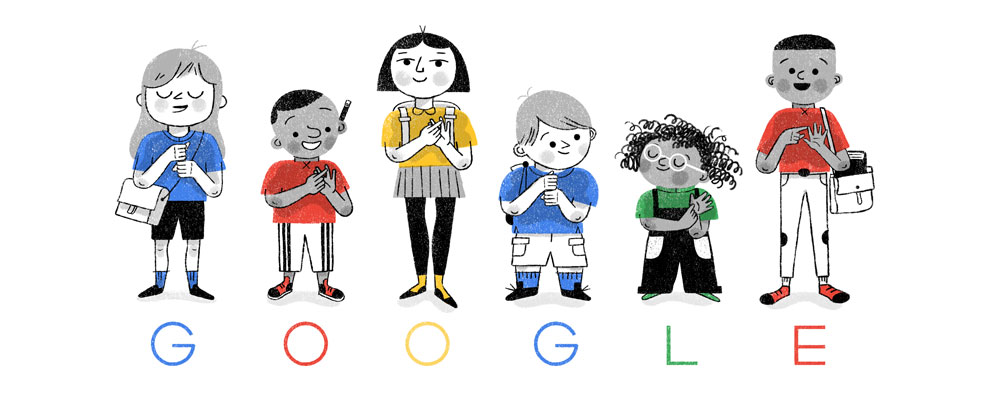}
\caption{A doodle from Google for celebrating sign language and the Braidwood Academy \cite{Google}. Avatars show the finger-spelling of the word \textit{Google}.}
\label{Doodle}
\centering
\end{figure}

Sign languages are languages that employ manual body gestures for communication. Signs are the alphabets in a sign language. A sign can be very intuitive and official, or it might take advantage of its visual nature for expressive or artistic effect, the words, the clauses, the pauses, the melody, the stressing or deemphasizing, the facial and vocal emotion, the body posture, head, hand gestures and others.

Different countries have different sign languages. Sign languages have been developed instinctively out of groups of individuals communicating with each other. Since sign languages have grown within the deaf societies, they can be segregated from the surrounding spoken languages. For example, despite the fact that English is the spoken language of England and United States, the American Sign Language (ASL) is quite different from British Sign Language (BSL). For a more detailed example, delivering a message through performing it with signs and gestures in American English environment, at least 3 approaches, called ASL, PSE (Pidgin Signed English) and Signed Exact English (SEE) are available. 
For instance, \textit{I went to the store yesterday} are performed by signs for ASL, PSE and SEE as \textit{Yesterday I go store}, \textit{I go store yesterday} and \textit{I go+ed to the store yesterday}, respectively.

In Fig. \ref{Doodle}, a doodle of Google is shown in which avatars show the finger-spelling of the word \textit{Google}. Common parameters of a sign language could be summarized as follows (Examples are reported based on ASL).

\begin{itemize}
\item \noindent \textbf{Palm orientation} Palm could face up, down, left, right, out and in. For example, consider \textit{baby} vs. \textit{table}.
\item \textbf{Hand Shape} Shape of hands and fingers are so useful for alphabetical letters and numbers. For example, consider \textit{I am Rita} vs. \textit{My Rita}. 
\item \textbf{Facial Expressions} Head nodes, head shakes, eyebrows, nose, eyes, lips, and emotions can be attached to the sign and bring a new meaning. For example, consider \textit{you} vs. \textit{is it you?} whose difference is in the surprising form of the eyes and eyebrows. Similarly, the hand shape and movement for \textit{I'm late} and \textit{I haven't} are same and just the face shape makes them different.  
\item \textbf{Location} Begin and end the sign at the correct position. Usually, signs originate from the body and terminate away or originate away from the body and terminate close to the body. For example, consider \textit{I'll see you tomorrow.}  
\item \textbf{Movement} Different kinds of movements are usually arc, straight line, circle, alternating in and out, the twist of the wrist and finger flick. In addition, in movement duration, the location, direction and also shape of the hands could change. For example, consider \textit{happy} or \textit{enjoy}. 
\end{itemize}

\begin{figure}[t]
\centering
\includegraphics[width=0.3\textwidth]{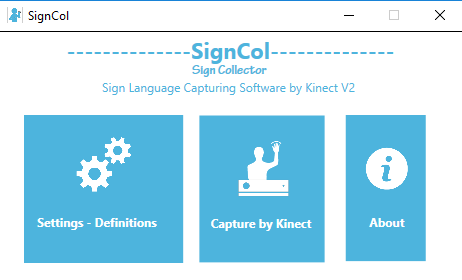}
\caption{The opening UI of the SignCol.}
\label{fig:opening}
\end{figure}

\subsection{Microsoft Kinect One (v2)}
Starting in 2013, Kinect for Xbox One (also called v2) was released with the Xbox One platform. Users need to acquire an adapter connecting to the Personal Computer (PC) in order to use the functionally of Kinect on PCs or laptops. V2 uses a wide-angle time-of-flight camera and processes 2 gigabytes of data per second to read the surrounding environment. Some of the embedded features can be summarized as Twenty five individual joints (including thumbs), 60 fps color camera captures 1080p video, monochrome depth-sensing video stream, which is in VGA resolution ($512 \times 424$ pixels) with 11-bit depth, active IR camera  $512 \times 424$ and 4 microphones. Moreover, the Software Development Kit (SDK) v2 can provide the location of the body, joint, and skeleton of bodies and it can track up to 6 bodies along with many other functionalities which are out pf scope for this paper. These capacities, enable Kinect to be used in areas far beyond a game console.  

\begin{figure}[]
\centering
\includegraphics[width=0.3\textwidth]{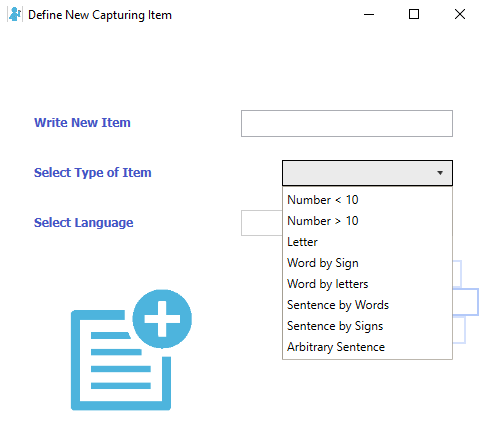}
\caption{The new item definition window of the SignCol.}
\label{fig:new_item}
\end{figure}

\subsection{Related work}
\label{related}
Sign language recognition (in a general context known as gesture recognition) is a complicated task and has gained a lot of attention in assistive technologies, computer vision, and human-machine interaction fields. In this section related literature and software are mentioned. 


The most similar software to SignCol is Complete Viewer, implemented in C++ and developed by Telecommunication System Engineering, Marche Polytechnic University~\cite{ItalyKinect, cippitelli2015kinect}. Complete Viewer is used to save the streams provided by Kinect V2. The saved streams include depth, RGB, mapping matrix, Infrared, Skeleton and time-stamp information. Signcol captures all of these streams along with providing database management for sign language items. In addition, SignCol saves body index frames and coordinate mapped bodies. 

Kinect2Viewer is a simple viewer sample for Kinect v2 written in C\#~\cite{Kinect2Viewer}. Kinect2Viewer illustrates and saves the stream data retrieved from Kinect v2 including color, depth, and skeleton. 


GesturePak records gestures and its SDK for WPF (.NET 4.5) determines when a user has made those gestures. GesturePak saves gestures including skeleton and body joints in XML format~\cite{GesturePak}.

\section{Design and implementation}
\label{Main}
\subsection{Overview}
\label{overview}
SignCol is developed to capture and store visual data from Microsoft Kinect v2, manage the database and show the statistics of the captured signs. The user in SignCol can define the following properties: language, capturing items and their categories, performers and capture the items while watching the statistics of the type of captured items to distribute the database species. SignCol can also handle signs with more than a person in the scene. 

As mentioned in section \ref{sign-language}, sign language is a complicated environment with different performing types, styles and constraints. To cover this environment effectively, we designed eight categories for the type of capturing items:
\begin{itemize}
\item cat1 -- Number $<$ 10 -- such as '4', '8'
\item cat2 -- Number $>$ 10 -- such as '16', '222'
\item cat3 -- Alphabet Letter -- such as 'A', 'F'
\item cat4 -- Word by a Sign -- such as 'I', 'My', 'Mom'
\item cat5 -- Word by Letters -- such as 'Entropy', 'Fourier'
\item cat6 -- Sentence by Words (by concatenated letters) -- such as 'Entropy learning', 'Fourier Transform'
\item cat7 -- Sentence by Signs -- such as 'I love you'
\item cat8 -- Arbitrary sentence -- such as 'Entropy of Mike's image is high'
\end{itemize}

\begin{figure}[t]
\centering
\includegraphics[width=0.5\textwidth]{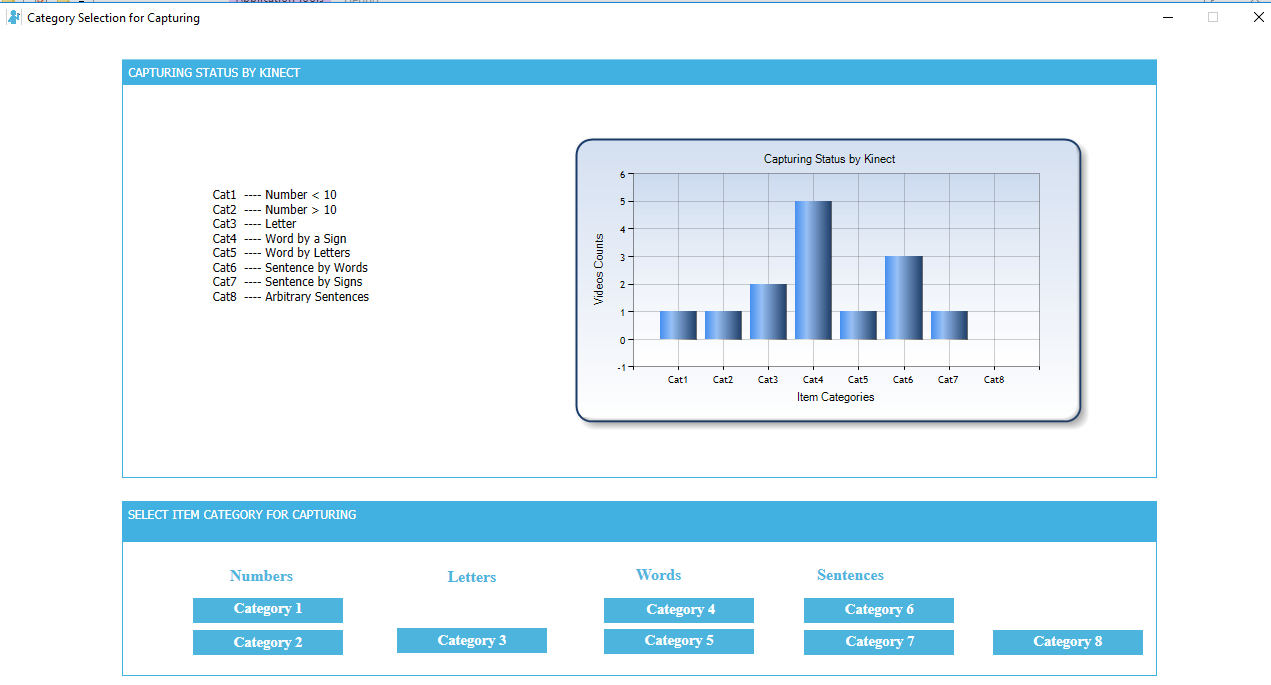}
\caption{Capturing window of the SignCol with the chart of  statistics and category selection.}
\label{fig:caputring}
\end{figure}

In an arbitrary category, both letters and signs are used. Notice that for many sentences, there is no strict category and they can be performed in different fashions. For example, \textit{I hate that} can be performed both in cat6 as concatenated words or as concatenated signs.

The data for each capturing record includes depth camera intrinsic parameters, time information and skeleton information of each frame, 8bit color frames, 16 bit depth frames, 8 bit infrared frames, 8 bit body index frames and frames of the coordinate mapped of color bodies which all of them are saved on hard disk in a folder created with following policy \\ \textit{Language\_Category\_Item\_Performer\_RandonNumber}. Frames are saved as a set of numbered images. Each type of above-mentioned data are located in a corresponding folder (e.g. depth frames folder is called \textit{depth frames}). The camera parameters are saved in text files. The time and skeleton information are saved in separate CSV files in different folders. In skeleton folder, for each frame there is a CSV file containing 25 rows according to 25 joints of body and columns for specifying the joint type, locations and space environments. As shown in Table \ref{skel-table}, each joint location is saved in three spaces, 3D camera space, 2D depth space (image) and 2D color space (image).

\begin{table*}[t]
\centering
\caption{An example of the format of the skeleton information in CSV file.}
\label{skel-table}
\begin{tabular}{cccccccc}
\textbf{JointType:} & WristRight & \textbf{CameraSpacePoint:} & X:0.0 Y:0.0 Z:0.0 & \textbf{DepthSpacePoint:} &  X:0.0 Y:0.0          & \textbf{ColorSpacePoint:} &      X:0.0 Y:0.0        \\
\textbf{JointType:} & HandRight  & \textbf{CameraSpacePoint:} & X:0.0 Y:0.0 Z:0.0               & \textbf{DepthSpacePoint:} &  X:0.0 Y:0.0          &   \textbf{ColorSpacePoint:} &    X:0.0 Y:0.0         
\end{tabular}
\end{table*}

\begin{figure}[t]
\centering
\includegraphics[width=0.45\textwidth]{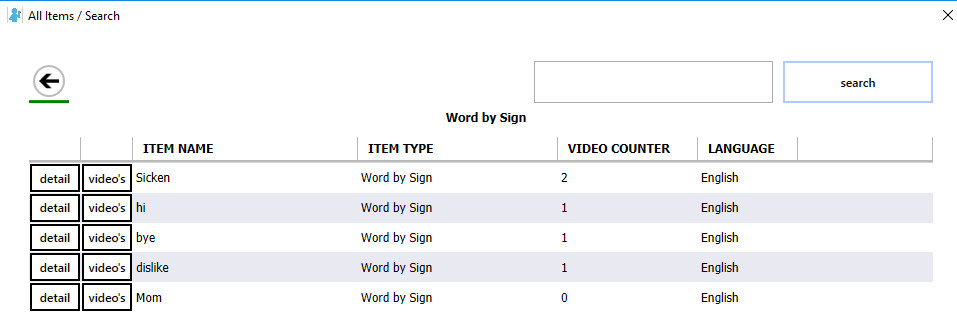}
\caption{List of items in a category with number of videos.}
\label{fig:items_cat}
\end{figure}

\subsection{User Interface and Forms}
\label{UI}

In this section, we provide descriptions over SignCol's user interface (UI). \cite{video1} is a video illustrating SignCol components and how to use it. The opening form is shown in figure~\ref{fig:opening}. The form contains buttons for three options, adjusting settings  and definitions, capturing images by Kinect and about. The settings form contains buttons linking to other auxiliary forms for defining language, items, saving path, and showing the list of performers, items and statistic charts. Figure~\ref{fig:new_item} shows the form of new item definition which has the fields for name, type of 8 category, and the corresponding language. 


By clicking the \textit{Capture by Kinect} button on the opening form, the form shown in figure~\ref{fig:caputring} will be displayed. This form has the chart about the statistics of the captured items with regards to the categories. It also contains button links to capture in each category. In fact, by considering the chart, the user will decide to capture an item from which category to make a uniform distribution of visual data. By clicking on a category, the form enlisting the items in that category is shown. This list also includes the number of captured videos regarding each item. The form also has the search option and is shown in figure~\ref{fig:items_cat}. By clicking on the videos button, the video list page will be opened and the user can select new capture button. 

\begin{figure}[t]
\centering
\includegraphics[width=0.48\textwidth]{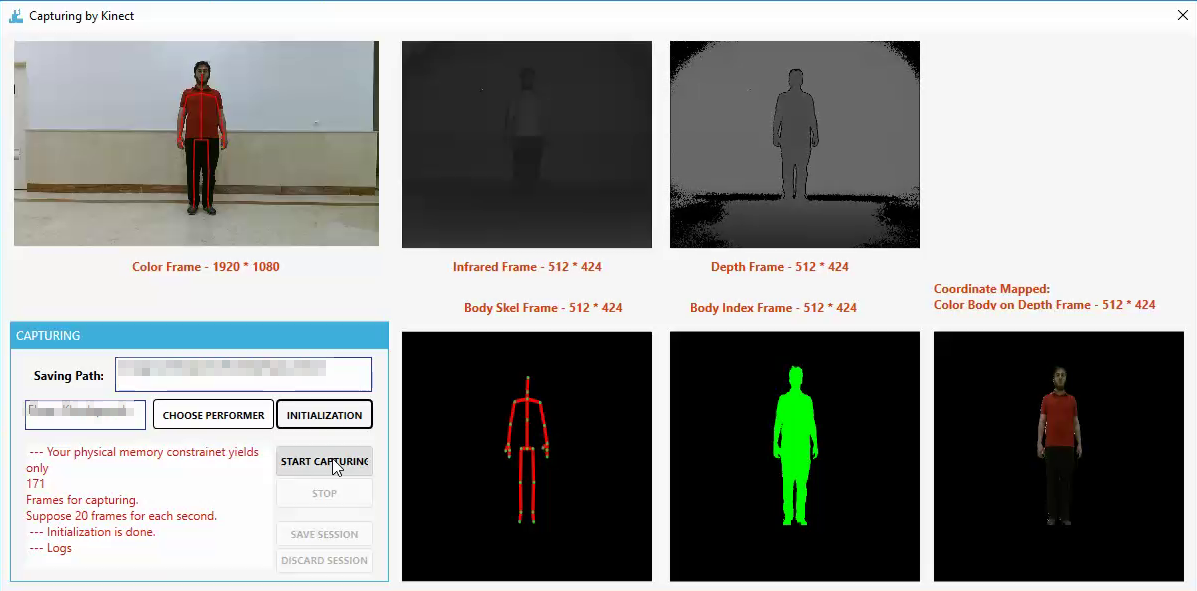}
\caption{The capturing form of the SignCol includes real time frames of RGB, infrared, depth, skeleton, body index, coordinate mapped bodies and buttons for actions.}
\label{fig:kinect}
\end{figure}

The capturing form is illustrated in figure~\ref{fig:kinect}. In this form, the user can select performer, initialize/start/stop capturing and save or discard the session and data.

\subsection{Implementation}
SignCol is an open source software available at~\cite{github} and is developed by Visual Studio Community 2015 based on c\#, Windows Presentation Foundation (WPF) graphical subsystem along with Model View View Model (MVVM) architecture~\cite{hall2011pro}. The database is generated and managed by SQLite engine~\cite{sqlite1}. The database contains tables for options, capturing items, languages, videos, and performers. Language table consists of defined language IDs. Name, age, and phone number are the fields of the performer table. Item table includes item id, corresponding language id, type id and its name. The video table contains video id, folder path, performer id, and item id. The location of the database is in the \textit{App\_Data} of the project. The database could be edited manually by using \textit{DB Browser for SQLite}.    

Followings are the requirements for a PC for compatibility with the Kinect v2 and SignCol: $\ast$ An Intel based motherboard (No AMD!) and Intel USB 3.0 chipset, $\ast$ 64-bit (x64), $\ast$ Physical core i5, 2.8 GHz or faster, $\ast$ USB 3.0 controller dedicated to the Kinect for Windows v2 sensor, $\ast$ 8 GB of RAM, $\ast$ Graphics card that supports DirectX 11, $\ast$ Windows 8, 8.1, 10, $\ast$ Kinect for Windows SDK v2.0, $\ast$ Kinect v2 (Kinect one) and it's PC-adapter and $\ast$ LCD resolution $1280 \times 800$ or above.

\section{Conclusion}
\label{conclusion}
This paper introduces an open source software called SignCol for collecting and storing data required for sign language recognition. SignCol is versatile in terms of letting user define languages, item categories and performers. It also provides users with statistics of the stored items. SignCol saves 6 different data modalities of gestures, such as RGB Color frames and depth frames. Extending the software to support multi-device multi-view visual capturing is an open task for future work.

\section*{Acknowledgement}
We gratefully thank Ehsan Khodapanah, Mohammad Reza Riahi and Dr. Hamid Safavi for their supports and testing efforts.  


\bibliography{refs}
\bibliographystyle{IEEEtran}

\section{Appendix}

In this section, we provide more screen shots of the software. \\

\begin{figure}[h]
\centering
\includegraphics[width=0.3\textwidth]{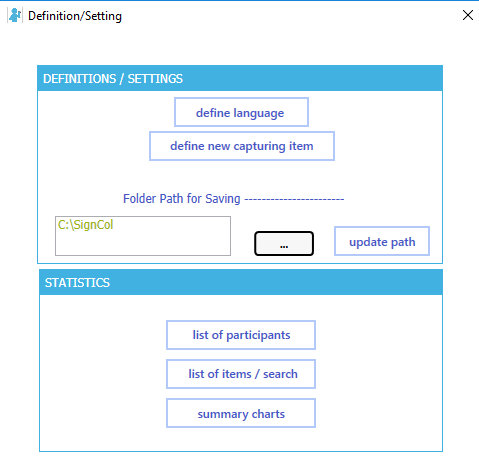}
\caption{Illustration of \textit{Settings-Definitions} form.}
\end{figure}

\begin{figure}[h]
\centering
\includegraphics[width=0.45\textwidth]{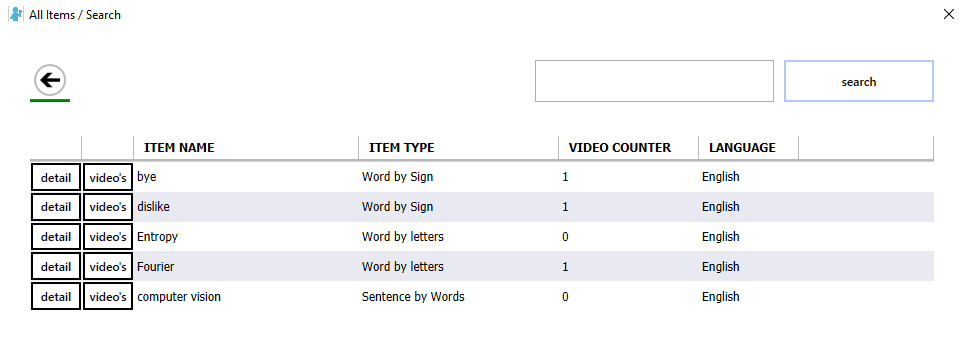}
\caption{Illustration of the \textit{items} form.}
\end{figure}

\begin{figure}[h]
\centering
\includegraphics[width=0.45\textwidth]{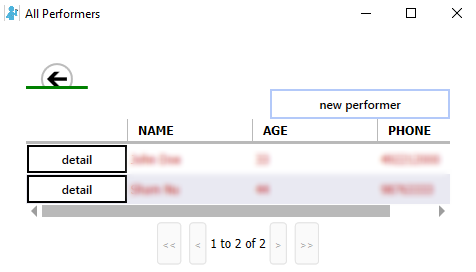}
\caption{Illustration of the \textit{participants} form.}
\end{figure}

\begin{figure}[h]
\centering
\includegraphics[width=0.45\textwidth]{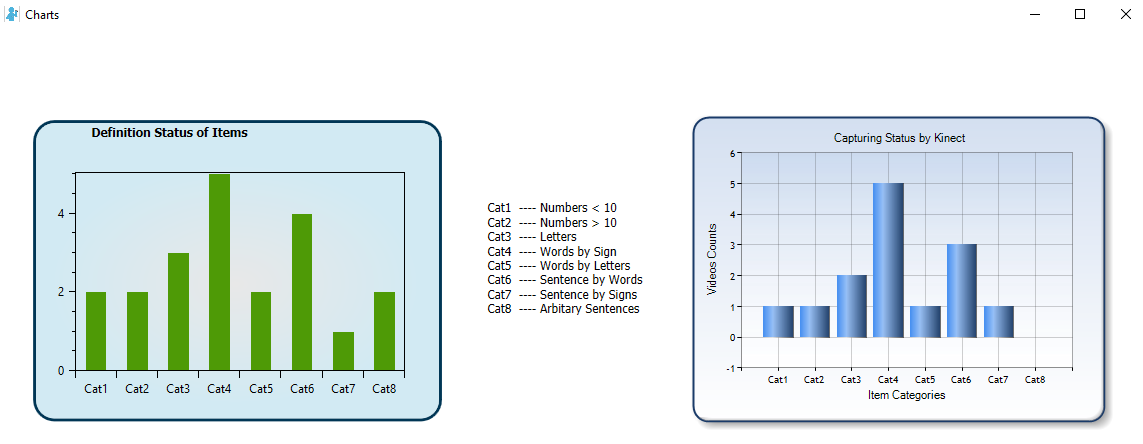}
\caption{Illustration of the \textit{Summary charts} form.}
\end{figure}

\begin{figure}[h]
\centering
\includegraphics[width=0.45\textwidth]{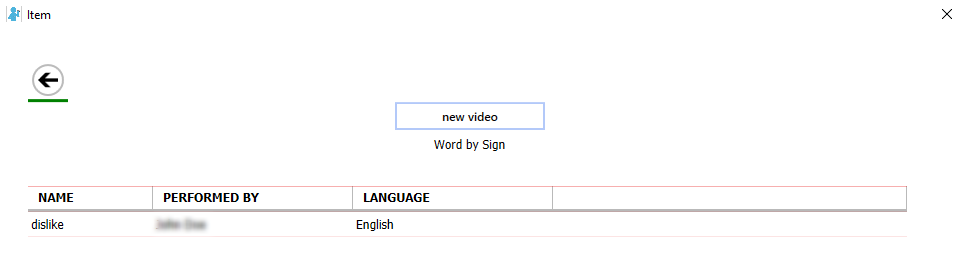}
\caption{Illustration of the \textit{item} form for capturing.}
\end{figure}

\begin{figure}[h]
\centering
\includegraphics[width=0.45\textwidth]{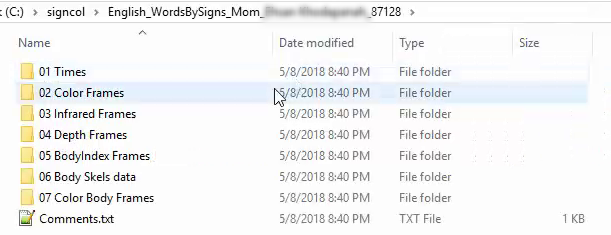}
\caption{Illustration of the folder includes the saved images and information.}
\end{figure}

\begin{figure}[h]
\centering
\includegraphics[width=0.45\textwidth]{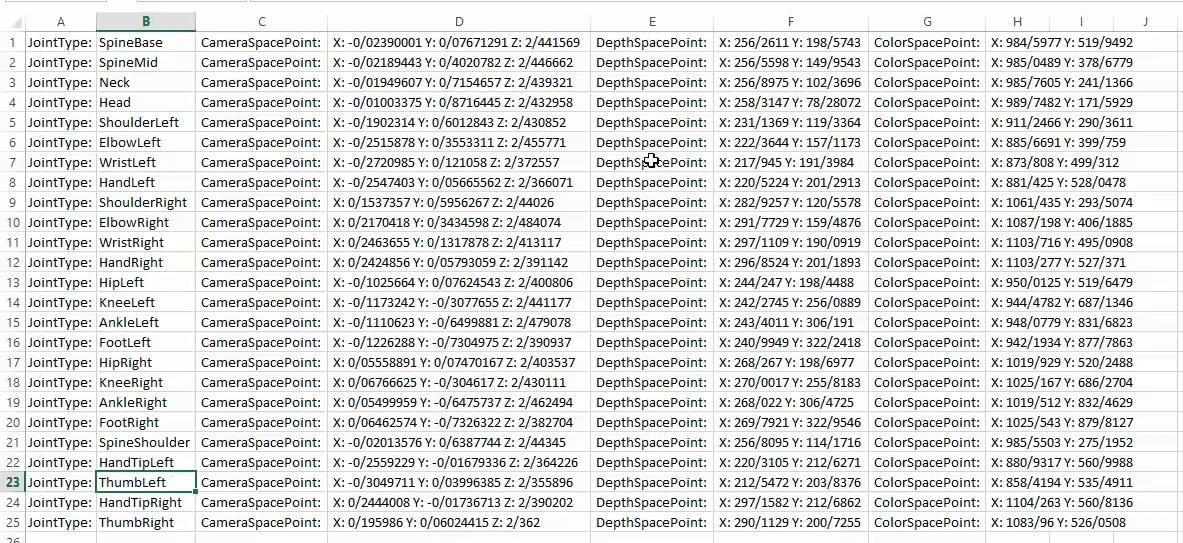}
\caption{Illustration of the skeleton information of a frame in a CSV file.}
\end{figure}


%

\end{document}